\documentclass[prc,aps,amsfonts,amsmath,superscriptaddress,floatfix,twocolumn]{revtex4}
\usepackage{graphicx,float} 
\usepackage{epsfig} 
\usepackage{rotating}
\usepackage{mwe}    
\usepackage{subfig}

\begin{document}

\title{Microscopic description of the self-conjugate $^{108}$Xe and $^{104}$Te $\alpha$-decay chain}
\author{F. Mercier}
\affiliation{IJCLab, Universit\'e Paris-Saclay, CNRS/IN2P3, 91405 Orsay Cedex, France}
\author{J. Zhao}
\affiliation{Center for Quantum Computing, Peng Cheng Laboratory, Shenzhen 518055, China}
\author{ R.-D Lasseri}
\affiliation{ESNT, CEA, IRFU, D\'epartement de Physique Nucl\'eaire, Universit\'e Paris-Saclay, F-91191 Gif-sur-Yvette}
\author{J.-P. Ebran}
\affiliation{CEA,DAM,DIF, F-91297 Arpajon, France}
\author{E. Khan}
\affiliation{IJCLab, Universit\'e Paris-Saclay, CNRS/IN2P3, 91405 Orsay Cedex, France}
\author{T. Nik\v si\' c}
\affiliation{Physics Department, Faculty of Science, University of Zagreb, 10000 Zagreb, Croatia}
\author{D. Vretenar}
\affiliation{Physics Department, Faculty of Science, University of Zagreb, 10000 Zagreb, Croatia}

\begin{abstract} 
A microscopic calculation of half-lives for the recently observed $^{108}$Xe $\to$ $^{104}$Te $\to$ $^{100}$Sn $\alpha$-decay chain is performed using a self-consistent framework based on energy density functionals. The relativistic density functional DD-PC1 and a separable pairing interaction of finite range are used to compute axially-symmetric deformation energy surfaces of $^{104}$Te and $^{108}$Xe as functions of quadrupole, octupole and hexadecupole collective coordinates. Dynamic least-action paths are determined that trace the $\alpha$-particle emission from the equilibrium deformation to the point of scission. 
The calculated half-lives: 197 ns for $^{104}$Te and 50 $\mu$s for $^{108}$Xe, are compared to recent experimental values 
of the half-lives of superallowed $\alpha$-decay of $^{104}$Te: $< 18$ ns, and $^{108}$Xe: 58$^{+106}_{-23}$ $\mu$s. 
\end{abstract}
 
\date{\today}

\maketitle

A fully microscopic description of $\alpha$-radioactivity presents a complex and difficult quantum mechanical problem. A  number of models of the $\alpha$-decay process have been developed over the past century, from semi-classical approaches to microscopic ones. In fact, one of the first successful models was the well-known Geiger-Nuttall law \cite{gei11} and its interpretation involving the tunneling effect by Gamow \cite{gam28}, and the class of WKB models \cite{del10}. From the experimental point of view, even the identification of all $\alpha$-emitting nuclei may have not been achieved yet, as demonstrated by the discovery of $\alpha$-emission from $^{209}$Bi in 2003 \cite{mar03}, or the remaining question of possible $\alpha$-radioactivity of $^{208}$Pb \cite{bee13}.

Several recent studies have been devoted to a more microscopic description of this process, such as the $\alpha$ pre-formation factor obtained form single-particle states calculated from a complex-energy shell-model \cite{bet12}. Microscopic-macroscopic approaches based on Woods-Saxon potentials have also been developed that consider an additional pocket-like surface potential \cite{bar16}, or a least-energy trajectory to describe $\alpha$ and cluster emission, and fission of $^{222}$Ra \cite{mir17}. Among fully microscopic approaches, we note the description of cluster emission from heavy nuclei using a self-consistent Hartree-Fock-Bogoliubov method based on the Gogny energy density functional (EDF) \cite{war11,rob18}.

Nuclear energy density functionals and, in particular, relativistic EDFs provide a natural framework for $\alpha$-decay studies.  They have been used to successfully describe the formation of $\alpha$-cluster states in light nuclei \cite{ebr12,ebr13,ebr14,ebr14a,zha15a,zho16}, including a quantitative comparison with experimental spectra of cluster states  in Ne isotopes \cite{mar18}, and the Hoyle state in $^{12}$C \cite{mar19}. Therefore, it could be interesting to study $\alpha$-radioactivity using models based on relativistic EDFs. On the one hand this approach can describe $\alpha$-cluster formation, and its qualitative relation with $\alpha$-emission \cite{ebr18}. On the other, it has already been successfully applied to spontaneous and induced fission dynamics \cite{zha15,zha16,tao17,zha19,zha19a}.  

The present study of $\alpha$-radioactivity is focused on the region of self-conjugate nuclei northeast of $^{100}$Sn. It is the lightest region of the nuclear mass table in which $\alpha$-particle emission has been identified, with the recent determination of the half-lives of superallowed $\alpha$-decay of $^{108}$Xe: 58$^{+106}_{-23}$ $\mu$s, and $^{104}$Te: $< 18$ ns \cite{aur18} (see also Ref.~\cite{xia19}). From a conceptual point of view, the lighter the nucleus the more localised the nucleonic wave-functions, an effect that arises both from the single-nucleon potential and the radial quantum numbers \cite{ebr12,ebr13,ebr18}. Hence the $^{100}$Sn region that includes the lightest $\alpha$-emitters presents the best case for  a microscopic approach based on the least-action integral on the potential energy surface (PES). In this work we will compute $\alpha$-decay half-lives of $^{108}$Xe and  $^{104}$Te using a model based on relativistic EDFs. 

The RHB framework is described in Refs.~\cite{vre05,men16,ebr19}. It provides a unified description of particle-hole ($ph$) and particle-particle ($pp$) correlations by combining two average potentials. Self-consistent calculations of deformation energy surfaces are performed using the DD-PC1~\cite{nik08} relativistic functional. In addition to the particle-hole channel determined by the choice of the EDF, a separable pairing interaction of finite-range \cite{dug04,tia09} is used that reproduces the pairing gap in nuclear matter as calculated with the D1S parametrization of the Gogny force~\cite{ber91,tia09}. The PES are calculated using the quadrupole, octupole and hexadecupole deformations as collective degrees of freedom. Deformation-constrained calculations are performed using a method with linear constraints that has successfully been applied to fission (see Ref.~\cite{zha15} for details). The Dirac-Hartree-Bogoliubov equations are solved by expanding the nucleon spinors in the basis of a 3D harmonic oscillator. Since $^{108}$Xe and  $^{104}$Te are not particularly heavy nuclei, calculations have been performed in a basis with 16 oscillators shells.

The process of emission of an $\alpha$-particle is modelled along a path $L$, determined by minimizing the action integral~\cite{bra72,led73}:

\begin{equation}
\label{eq:act_integration}
S(L) = \int_{s_{\rm in}}^{s_{\rm out}} {1\over\hbar} 
  \sqrt{ 2\mathcal{M}_{\rm eff}(s) \left[ V_{\rm eff}(s)-E_0 \right] } ds ,
\end{equation}
where $\mathcal{M}_{\rm eff}(s)$ and $V_{\rm eff}(s)$ are the effective 
collective inertia and potential, respectively.
$E_0$ is the collective ground-state energy, and the integration
limits correspond to the classical inner ($s_{\rm in}$) and outer turning points ($s_{\rm out})$, defined by $V_{\rm eff}(s) = E_0$.

The effective inertia is computed from the multidimensional collective inertia tensor $\mathcal{M}$ \cite{bra72,bar78,bar81,sad13,sad14}
\begin{equation}
\mathcal{M}_{\rm eff}(s) = \sum_{ij} \mathcal{M}_{ij} {dq_i \over ds} {dq_j \over ds}\;,
\end{equation}
where $q_i(s)$ denotes the collective coordinate as a function of the path's length. The collective inertia tensor is calculated using the self-consistent RHB solutions and applying the adiabatic time-dependent Hartree-Fock-Bogoliubov (ATDHFB)  method \cite{bar11}. In the perturbative cranking approximation  
 the collective inertia reads \cite{zha15}

 \begin{equation}
\label{eq:pmass}
\mathcal{M} = \hbar^2 {\it M}_{(1)}^{-1} {\it M}_{(3)} {\it M}_{(1)}^{-1}, 
\end{equation}
 
 where 
\begin{equation}
\label{eq:mmatrix}
\left[ {\it M}_{(k)} \right]_{ij} = \sum_{\mu\nu} 
    {\left\langle 0 \left| \hat{Q}_i \right| \mu\nu \right\rangle
     \left\langle \mu\nu \left| \hat{Q}_j \right| 0 \right\rangle
     \over (E_\mu + E_\nu)^k}\; . 
\end{equation}
$|\mu\nu\rangle$ are two-quasiparticle 
wave functions, and $E_\mu$ and $E_\nu$ the corresponding quasiparticle energies. $\hat{Q}_i$
denotes the multipole operators that describe the collective degrees of freedom.
The effective collective potential $V_{\rm eff}$ is obtained by subtracting the 
vibrational zero-point energy (ZPE) from the total RHB 
deformation energy. Following the prescription of 
Refs.~\cite{sta13,bar07,sad13,sad14}, 
the ZPE is computed using the Gaussian overlap approximation, 
\begin{equation}
\label{eq:zpe}
E_{\rm ZPE} = {1\over4} {\rm Tr} \left[ {\it M}_{(2)}^{-1} {\it M}_{(1)} \right].
\end{equation}
The microscopic self-consistent solutions of the constrained RHB equations, that is, the 
single-quasiparticle energies and wave functions on the entire energy surface as functions of the 
quadrupole and octupole deformations, provide the microscopic input for the calculation of both the 
collective inertia and zero-point energy. 

In practical calculations we first determine the least action path from ground state to scission in the restricted 2-dimensional 
collective space ($\beta_{20},\beta_{30})$. The scission point is determined by a discontinuity in $\beta_{40}$.
After scission, the configuration with two well separated fragments becomes the lowest energy solution and the energy 
can be approximated by the classical expression for two uniformly charged spheres: 
\begin{equation}
\label{eq:coulomb}
V_{\rm eff}(\beta_{3}) = e^2 {Z_{1}Z_{2} \over R} - Q,
\end{equation}
where $R$ represents the distance between the centers of mass of the fragments, and 
the second term is the experimental $Q$ value.
We use Eqs. (9) and (10) of Ref. \cite{war11} to approximate the relation between $R$ and the octuple moment $Q_{30}$,
\begin{equation}
\label{eq:R-Q}
Q_{30} = f_3 R^3,
\end{equation}
with
\begin{equation}
\label{eq:f3}
f_3 = {A_1 A_2 \over A} {(A_1 - A_2) \over A},
\end{equation} 
and $\beta_{30}=4\pi Q_{30} / 3AR^3$.
The corresponding effective collective mass reads
\begin{equation}
\label{eq:cmass}
\mathcal{M}_{\rm eff} = {\mu \over 9Q^{4/3}_{30} f^{2/3}_3}, 
\end{equation}
where $\mu=m_{n} A_1A_2/(A_1+A_2)$ is the reduced mass of the two fragments, and 
$m_{n}$ denotes the nucleon mass \cite{war11}. 
Thus the path involved in the action integral of Eq. (\ref{eq:act_integration}) consists of the least-action path 
from $s_{\rm in}$ to scission, and the energy is approximated by the Coulomb potential from scission to $s_{\rm out}$ \cite{war11}. 
The alpha decay half-life is calculated as $T_{1/2}=\ln2/(nP)$, where $n$ is the
number of assaults on the potential barrier per unit 
time~\cite{bar78,bar81,sad13,sad14}, 
and $P$ is the barrier penetration probability in the WKB approximation
\begin{equation}
P = {1 \over 1+\exp[2S(L)]}. 
\label{prob}
\end{equation}
We choose $E_{0} = 1$ MeV in Eq. (\ref{eq:act_integration}) for the value of the collective ground state energy.
For the vibrational frequency $\hbar\omega=1$ MeV, the corresponding $n$ value is $10^{20.38}$ s$^{-1}$.

\begin{figure}[tbh!]
\scalebox{0.33}{\includegraphics{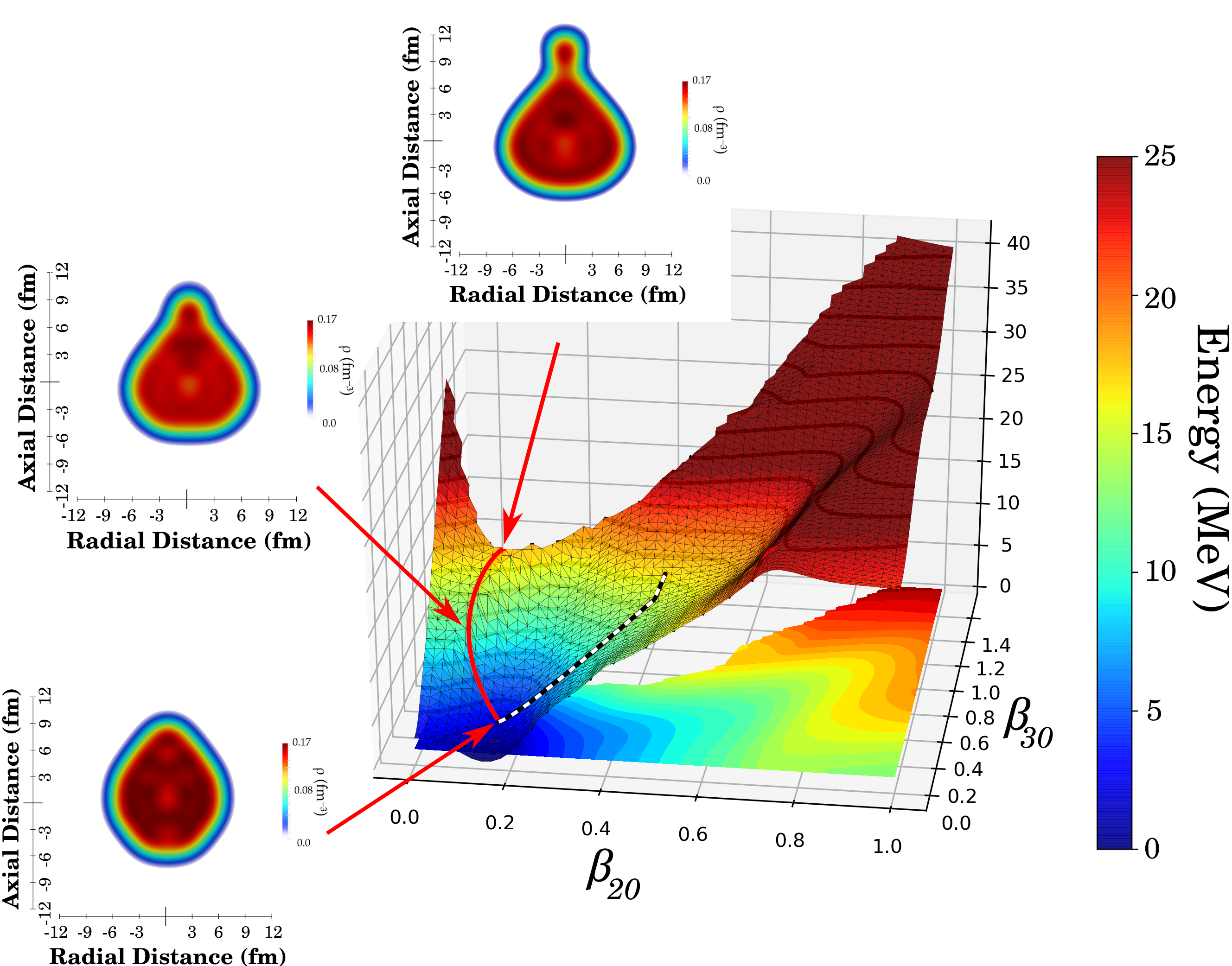}}
 \caption{Deformation energy surface of $^{104}$Te in the quadrupole-octupole axially-symmetric plane, calculated with the RHB model based on the DD-PC1 functional. The dashed and solid curves on the energy surface correspond to the static (least-energy) and dynamic (least-action) paths for $\alpha$-emission, respectively. The insets display the intrinsic nucleon densities at selected values of ($\beta_{20}, \beta_{30}$).}    
 \label{fig:104PES}
\end{figure}

Figure \ref{fig:104PES} displays the axially-symmetric deformation energy surface of $^{104}$Te with respect to the octupole and quadrupole collective coordinates. When one considers the paths for $\alpha$-emission, the static path (dashed) which minimizes the energy without taking into account the effect of the collective inertia evolves towards larger quadrupole deformation. More relevant is the dynamic path (solid), which explicitly takes into account the collective inertia by minimizing the action integral of Eq. (\ref{eq:act_integration}). The numerical minimization technique used to determine the dynamic path is described in Ref. \cite{zha15}. We have considered several possible values for the turning point $s_{\rm in}$ and the scission point to make certain that the minimum action path is chosen. When compared to the static path, the nucleus exhibits smaller quadrupole deformations along the dynamic path to $\alpha$-emission. 
The insets show how the total intrinsic nucleon densities evolve from the equilibrium deformation to the end point 
of the dynamic path, which corresponds to the scission of a small cluster of nucleons. The integral of the density of this 
cluster is close to 4 nucleons.

\begin{figure}[tbh!]
\scalebox{0.33}{\includegraphics{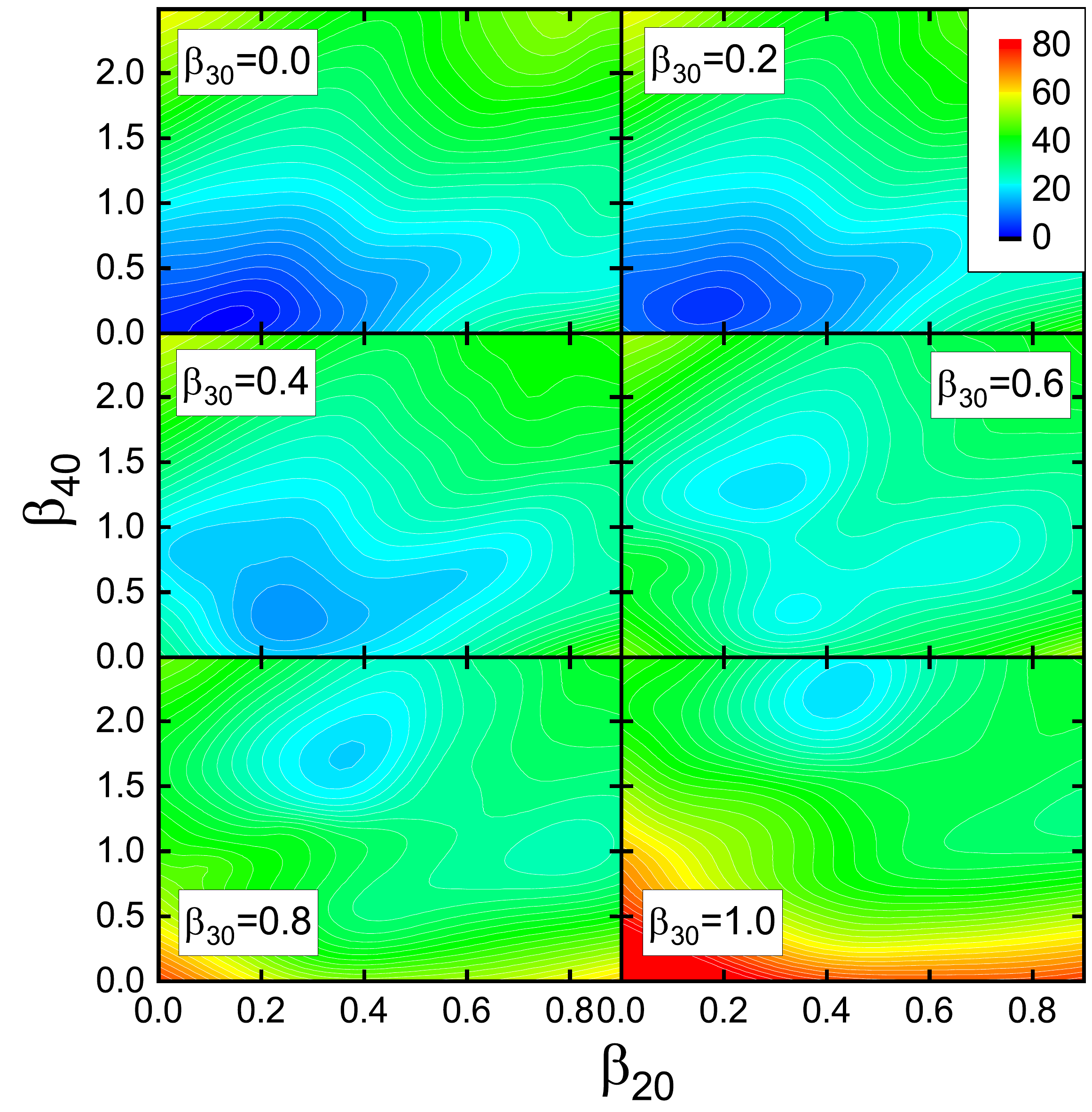}}
 \caption{Deformation energy surface of $^{104}$Te in the quadrupole-hexadecupole axially-symmetric plane, calculated with the RHB model based on the DD-PC1 functional, for selected values of the octupole deformation $\beta_{30}$. Contours join points on the surface with the same energy (in MeV).}    
 \label{b2b4}
\end{figure}

In Fig.~\ref{b2b4} we illustrate the corresponding evolution of hexadecupole deformation. The self-consistent deformation energy surface of $^{104}$Te is shown in the ($\beta_{20}, \beta_{40}$) plane, for selected values of the octupole deformation parameter $\beta_{30}$. For large octupole deformation ($\beta_{30} \gtrsim 0.45$), one notices a pronounced sudden increase of the $\beta_{40}$ value at the energy minimum (from typically 0.1 to about 1.5).  This jump corresponds to the scission between the $\alpha$-particle and the remaining $^{100}$Sn nucleus. After scission, as discussed in Ref. \cite{war11}, Coulomb repulsion between the two fragments determines the dynamics, 
and the corresponding collective potential and collective mass are therefore calculated using Eqs. (\ref{eq:coulomb}) - (\ref{eq:cmass}). The experimental $\alpha$-decay $Q$ value is taken from Ref. \cite{aur18}. The $\alpha$-decay half-life of $^{104}$Te calculated with Eq. (\ref{prob}) is 197 ns, and this result can be compared to the recent experimental value of $< 18$ ns \cite{aur18}. 

\begin{figure}[htb!]
\scalebox{0.31}{\includegraphics{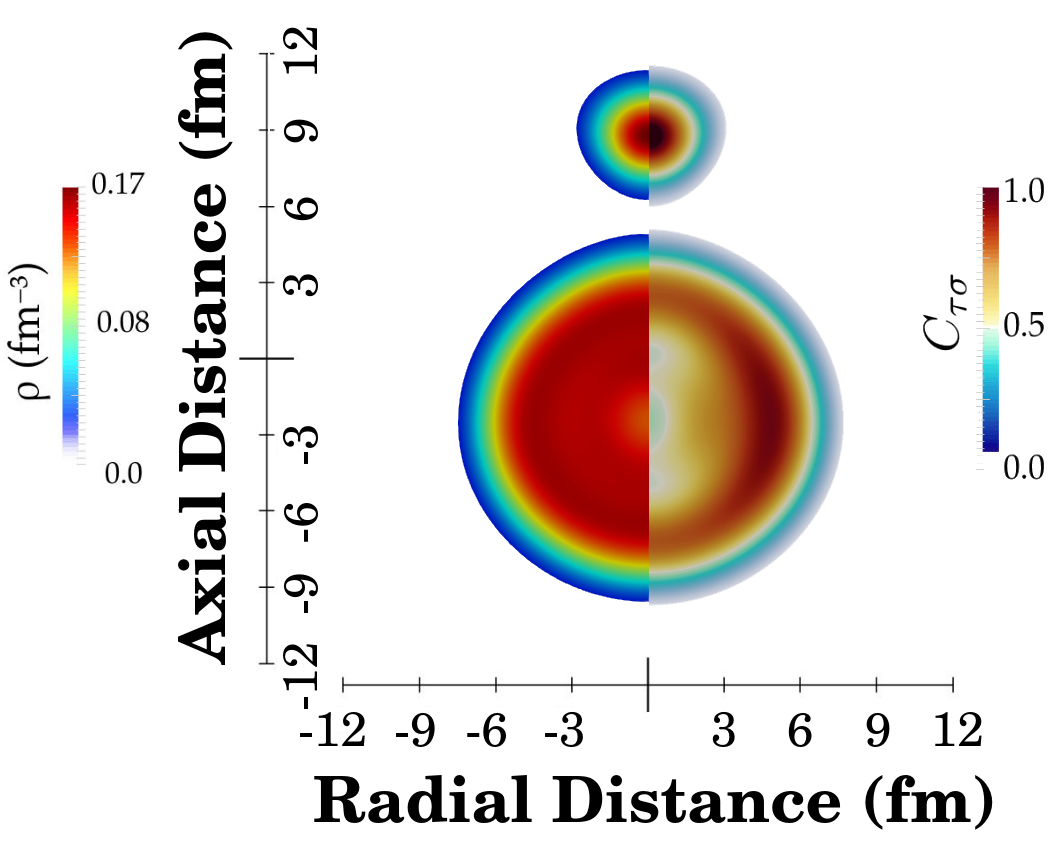}}
 \caption{Total nucleon density of the fragments around scission for $\alpha$-emission from $^{108}$Xe (left panel). The corresponding Fermion localization function Eq.~(\ref{localization}) is shown in the panel on the right.}    
 \label{fig:108Xedens}
\end{figure}

We have also performed a corresponding analysis of $\alpha$-decay of the next $N=Z$ nucleus $^{108}$Xe. The deformation energy surface in the ($\beta_{20}, \beta_{30}$) plane is shown in Fig. \ref{fig:108PES}, where we also include the 
dynamic (least-action) path for $\alpha$-emission together with the intrinsic nucleon densities at selected values of $\beta_{20}$ and $\beta_{30}$.  The left panel of Fig.~\ref{fig:108Xedens} displays the nucleon density of the fragments around scission for $\alpha$-emission from $^{108}$Xe. The number of nucleons obtained by integrating the density of the $\alpha$-like fragment is close to four, up to a few percents. In the panel on the right we plot the corresponding Fermion localization probability calculated using the  expression:

\begin{figure}[tbh!]
\scalebox{0.13}{\includegraphics{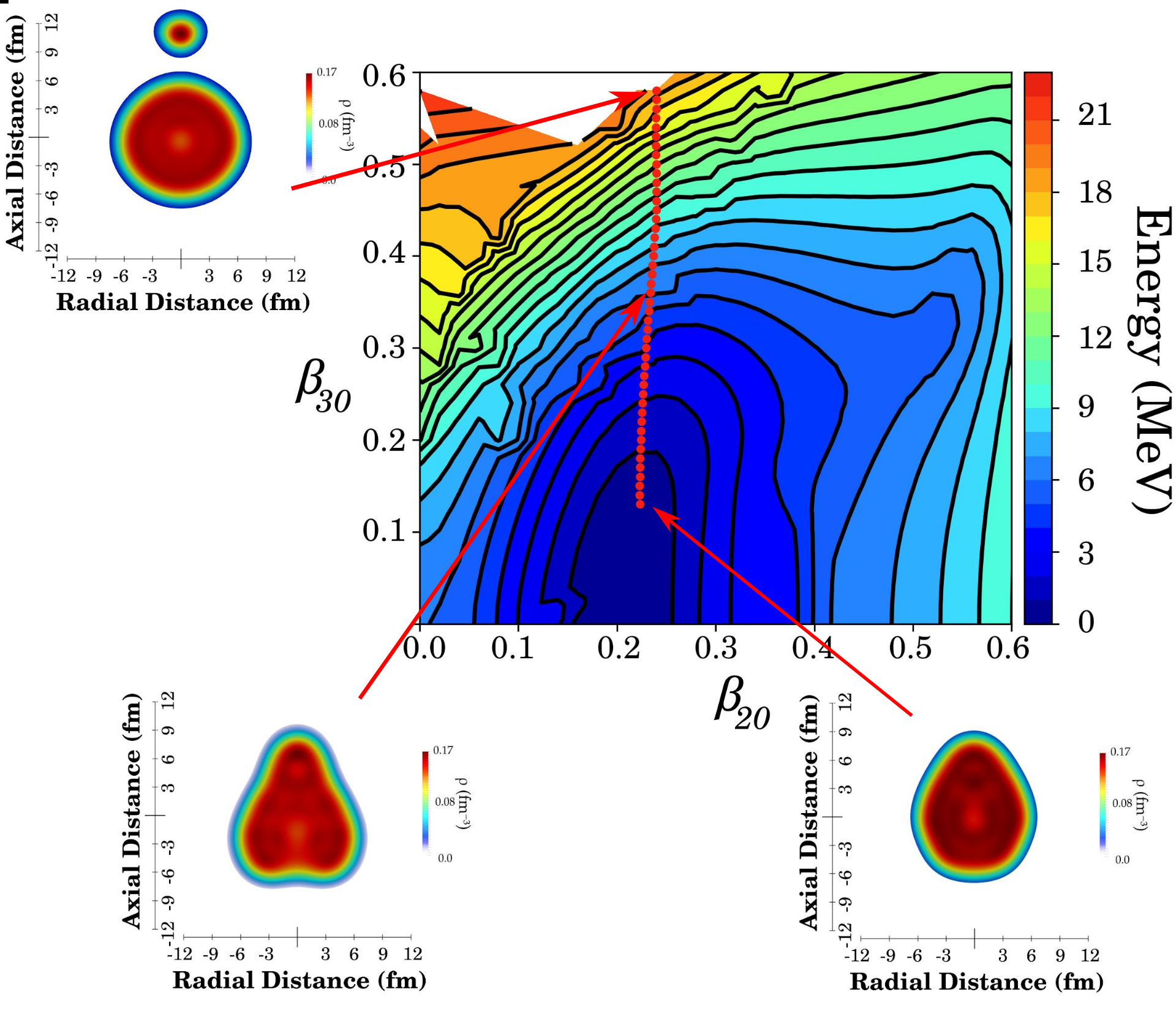}}
 \caption{Deformation energy surface of $^{108}$Xe in the quadrupole-octupole axially-symmetric plane, calculated with the RHB model based on the DD-PC1 functional. The contours join points on the surface with the same energy. The dynamic (least-action) path for $\alpha$-emission is shown together with intrinsic nucleon densities at selected values of ($\beta_{20}, \beta_{30}$).}    
 \label{fig:108PES}
\end{figure}

\begin{equation}
C_{q\sigma}(\vec{r})=\left[1+\left(\frac{\tau_{q\sigma}\rho_{q\sigma}-\frac{1}{4}\mid\vec{\nabla}\rho_{q\sigma}\mid^2-\vec{j}^2_{q\sigma}}{\rho_{q\sigma}\tau^{TF}_{q\sigma}}\right)^2\right]^{-1}
\label{localization}
\end{equation}
where $\rho_{q\sigma},\vec{j}_{q\sigma},\tau_{q\sigma},\tau^{TF}_{q\sigma}$  and $\vec{\nabla}\rho_{q\sigma}$ are the particle, current, kinetic, Thomas-Fermi kinetic densities and density gradient, respectively. A value close to 1 means that the probability of finding two nucleons with the same spin and isospin at the same point $\vec{r}$ is very small. This is the case for  alpha clusters because the four nucleons occupy different spin and isospin states and thus $C_{q\sigma}\simeq 1$ \cite{rei11,ebr17,jer18}. The fermion localization probability in Fig. \ref{fig:108Xedens} shows that the cluster emitted from $^{108}$Xe indeed corresponds to an $\alpha$-particle. The dynamic least-action path plotted in Fig. \ref{fig:108PES}, together with Eqs. (\ref{eq:act_integration}) and (\ref{prob}), is used to calculate the half-life for $\alpha$-decay: 50 $\mu$s, in close agreement with the experimental value: 58$^{+106}_{-23}$ $\mu$s \cite{aur18}.

In summary, the self-consistent mean-field framework based on relativistic energy density functionals, previously successfully applied to nuclear structure and fission dynamics, has been used to analyse the recently observed $\alpha$-decay chain 
$^{108}$Xe $\to$ $^{104}$Te $\to$ $^{100}$Sn. By employing the relativistic density functional DD-PC1 and a separable pairing interaction of finite range, axially-symmetric deformation energy surfaces in the plane of quadrupole and octupole collective coordinates have been mapped for $^{104}$Te and $^{108}$Xe by performing self-consistent constrained mean-field calculations. Dynamic least-action paths have been determined that allow to trace the process of $\alpha$-particle  emission from the mean-field equilibrium deformation to the point of scission. The latter is manifested by a sudden marked  increase of hexadecupole deformation, and after scission Coulomb repulsion between the two fragments determines the dynamics. By taking into account the collective inertia in the perturbative cranking approximation of ATDHFB, the resulting $\alpha$-decay half-lives of $^{104}$Te and $^{108}$Xe are in quantitative agreement with recently determined experimental values. 

\begin{acknowledgments} 
This work has been supported in part by the QuantiXLie Centre of Excellence, a project co-financed by the Croatian Government and European Union through the European Regional Development Fund - the Competitiveness and Cohesion Operational Programme (KK.01.1.1.01), and by the framework of the Espace de Structure et de r\'eactions Nucl\'eaires Th\'eorique (ESNT esnt.cea.fr) at CEA. J. Z. acknowledges support by the NSFC under Grant No. 11790325. F. Mercier and J. Zhao contributed equally to this work.
\end{acknowledgments}

\end{document}